\documentclass[10pt,conference,compsocconf]{IEEEtran}
\usepackage[utf8]{inputenc}
\usepackage[T1]{fontenc}
\usepackage{graphicx}
\usepackage{cite}
\usepackage{amssymb,amsmath}
\usepackage{url}
\usepackage[small]{caption}
\renewcommand{\dbltextfloatsep}{3pt}
\renewcommand{\textfloatsep}{3pt}
\renewcommand{\floatsep}{3pt}
\begin{document}

\newcommand{\superscript}[1]{\ensuremath{^{\textrm{#1}}}}
\def\sharedaffiliation{\end{tabular}\newline\begin{tabular}{c}}
\def\wu{\superscript{*}}
\def\wg{\superscript{\dag}}

\title{Navigating MazeMap: indoor human mobility, spatio-logical ties and future potential}

\author{
\IEEEauthorblockN{Gergely Bicz\'ok\IEEEauthorrefmark{1}, Santiago D\'iez Mart\'inez\IEEEauthorrefmark{1}, Thomas Jelle\IEEEauthorrefmark{1}\IEEEauthorrefmark{2}, and John Krogstie\IEEEauthorrefmark{1}}     
\IEEEauthorblockA{\IEEEauthorrefmark{1}Norwegian Univ. of Science and Technology, E-mail: \{gbiczok@item, santiagd@stud, krogstie@idi\}.ntnu.no}
\IEEEauthorblockA{\IEEEauthorrefmark{2}MazeMap, E-mail: thomas@mazemap.com} 
%{\IEEEauthorrefmark{1}  A@email.com}, {\IEEEauthorrefmark{2} B@email.com}, {\IEEEauthorrefmark{3} C@email.com}, \IEEEauthorrefmark{4} D@email.com}
}

\maketitle

\begin{abstract}
Global navigation systems and location-based services have found their way into our daily lives. Recently, indoor positioning techniques have also been proposed, and there are several live or trial systems already operating. In this paper, we present insights from MazeMap, the first live indoor/outdoor positioning and navigation system deployed at a large university campus in Norway. Our main contribution is a measurement case study; we show the spatial and temporal distribution of MazeMap geo-location and wayfinding requests, construct the aggregated human mobility map of the campus and find strong logical ties between different locations. On one hand, our findings are specific to the venue; on the other hand, the \emph{nature} of available data and insights coupled with our discussion on potential usage scenarios for indoor positioning and location-based services predict a successful future for these systems and applications.
\end{abstract}

\IEEEpeerreviewmaketitle

%\category{C.2.3}{Computer-Communication Networks}{Network Operations}
%\category{J.4}{Computer Applications}{Social and behavioral sciences}
%\terms{Measurement}

%\keywords{indoor navigation, indoor location-based services, measurement, campus}

\section{Introduction and Related Work}
\label{sec:intro}

Navigation systems have been incorporated to our daily life. The most prolific example is the Global Positioning System (GPS) used by motorists and pedestrians alike. Although GPS technology itself is quite mature, the market for devices with GPS capability is expected to grow at a Compound Annual Growth Rate of 15\%  between 2012 and 2015 \cite{gps_report}. Driving this steady growth is the increasing proliferation of smartphones with built-in GPS receivers enabling services beyond plain navigation such as location-based shopping and social networking. Despite the steady market expansion, location-based services would be crippled if limited only to outdoor venues: people in urban areas could spend around 80-90\% of their time indoors \cite{indoor_people}. Fortunately, indoor positioning technologies are on the rise, enabling smartphone users, businesses and software developers to determine the location of people and objects inside buildings. Most proposed Indoor Positioning Systems (IPS) utilize some form of wireless radio communications: WiFi, cellular, RFID, Bluetooth,  enhanced GPS and their respective combinations; please refer to \cite{ips_survey} for a starting point.
%, and to the website of the International Conference on Indoor Positioning and Indoor Navigation (IPIN) for fresh research results \cite{ipin_website}. 

IPS inherently has the potential to tremendously advance the concept of pervasive computing by forming the basis of services like indoor wayfinding, people and asset monitoring, personalized shopping, improved emergency response and even making the real world digitally searchable down to the object level \cite{indoor_forbes}. Given the intriguing business potential of an emerging multi-billion dollar market, it is hardly surprising that major tech companies are also involved in incorporating IPS into their platforms and products. On the hardware vendor side, Broadcom is focusing on enhancing the indoor positioning capabilities of their chips \cite{indoor_chip_broadcom}. Already present in the online map segment, Google \cite{indoor_map_google} and Nokia \cite{indoor_map_nokia} are working on extending their offerings for maps of indoor venues. Currently, there are hundreds of SMEs active in IPS-related business, and tech giants pay close attention as evidenced by Apple's \$20M acquisition of WifiSLAM \cite{indoor_apple}.

There is a long list of indoor venues, where users, businesses and venue-owners could mutually benefit from Indoor Location-Based Services (ILBSs) built on top of an IPS. This list includes shopping malls, large hotels and casinos, airports, hospitals, museums, university campuses and office buildings. Live and piloting shopping-related ILBSs include the ones offered by BestBuy \cite{indoor_shop_bestbuy} and other major retailers such as Macy's and Target, partnering with different IPS startups such as the already profitable Shopkick \cite{indoor_shop_shopkick}. Boston Children's Hospital has launched its MyWay mobile app providing among others indoor wayfinding \cite{indoor_venue_hospital}. In addition, the American Museum of Natural History provides visitors with a personalized, location-aware tour guide app \cite{indoor_personalized_museum}. 

NTNU (Norwegian University of Science and Technology) partnering with Wireless Trondheim has been offering a hybrid indoor/outdoor wayfinding app called MazeMap (earlier known as CampusGuide) \cite{campusguide,campusguide_press} for its main Gl\o shaugen campus since Fall 2011; the first of its kind. MazeMap is able to locate a user's position on campus with an accuracy of up to 5-10 meters, and provides room-level wayfinding and object search capabilities. In this paper we focus on MazeMap, and present a measurement study based on 19 month of user logs. Our main contribution is a first-of-its-kind case study providing usage statistics, human mobility patterns and a spatio-logical network of rooms from a live ILBS covering an entire university campus. First, we present spatial and temporal distribution of user requests at the building, room and object level. Second, we build a weighted directed graph out of turn-by-turn routes resulting from wayfinding requests. We construct a human mobility map, showing aggregated user mobility patterns and campus highways. Third, we consider the logical connections between rooms and floors linked by wayfinding requests. We show that there are expected and unexpected strong ties in this social graph. Furthermore, we find that there is strong correlation  between the strength of logical connections and geographical distance; in fact more than 70\% of wayfinding requests are intra-building. In addition, we give an outlook on the potential venues and usage scenarios for IPS such as shopping malls, hospitals, institution-level resource management, emergency preparedness and human mobility research.

One paper we are aware of which is close to our work is \cite{spatial:2013}; however, it is based on the experimental tracking of 37 users for one month, while our study spans thousands of users and 19 month in a large-scale, live system. Owing to our slightly different focus, we do not elaborate on wireless indoor positioning techniques; we only provide a starting point \cite{ips_survey}. We do provide a unique case study from a live ILBS and a short discussion on the potential of indoor positioning, however, our contribution has some limitations. Due to space limitation we do not provide a comprehensive analysis of the dataset: we believe that the \emph{nature} of insights provided by the case study are more important than the actual results. They justify the potential of ILBSs in multiple venues and scenarios. Moreover, the dataset is limited in that the system does not yet have the capability for tracking individual devices or user sessions; once operational, the latter feature is expected to provide deeper insights.

The rest of the paper is organized as follows. Section \ref{sec:MazeMap} gives an overview of the MazeMap system. Section \ref{sec:data} introduces the dataset and analysis methods. Section \ref{sec:results} presents our case study. Section \ref{sec:discussion} discusses potential future usage scenarios and challenges. Finally, Section \ref{sec:conclusion} concludes the paper.

%\pagebreak
\section{MazeMap}
\label{sec:MazeMap}

MazeMap, a service for indoor positioning and navigation, started as an R\&D  project between Wireless Trondheim and NTNU \cite{Krogstie:2012}. Each year, 5000 new students arrive at NTNU, and have lectures and activities all over NTNU's premises. The largest campus, Gl\o shaugen, covers 350000 sqm with over 60 buildings and 13000 rooms. It is therefore a big challenge for new students and many visitors to find their way around campus. This was the motivation for MazeMap; the first version was launched 31st August, 2011 under the name CampusGuide. MazeMap allows the user to see  building maps on campus, locate the user's own position within the building, search for all rooms and different objects (toilets, parking lots, etc.), and get turn-by-turn directions from where the users is to where he wants to go. MazeMap can be tested from anywhere at \cite{campusguide}. The service has become quite popular: during the start of a new semester nearly 10\% of all employees and students use the service daily; the total number of unique devices logged since August 2011 is around 20000. MazeMap aims to help users find their ways with their laptops, tablets and smartphones; the service is available from a browser and as an app at both Google Play and Apple AppStore.

MazeMap can locate a user's position indoors with an accuracy of up to 5-10 meters using the dense WiFi network of NTNU, which has more than 1,800 WiFi access points. The technique used is called trilateration \cite{trilateration}, where signal strengths from 3 or more different access points are measured, and the user's position is calculated based on these. The accuracy of the position depends on the WiFi network, how access points are placed and their density. At this specific campus, most outdoor areas are also covered multiple access points; however, MazeMap is also able to use standard GPS positioning when the user is outdoors (if the device has an enabled GPS receiver), providing approximately the same precision as the WiFi method. Note that the system is also able to combine WiFi and GPS signals if available. On the user interface side, in order to present the users with readable maps, the service uses construction drawings, and interprets them to recognize different objects, and to choose what to show or hide. This way the maps the users see are not overloaded with unnecessary technical data.

MazeMap is delivered as a software as a service, which enables frequent updates of design and new functionality. The service is already launched at different other premises including St. Olavs Hospital (regional university hospital) and the University of Troms\o; other venues are scheduled to start soon. Furthermore, as part of the Wireless Trondheim Living Lab \cite{Andresen:2007} there is an interdisciplinary group at Wireless Trondheim and NTNU looking into novel ways of utilizing location data and improving the navigation service. MazeMap therefore collects depersonalized data of service usage including positioning logs, in accordance with NTNU directives and Norwegian law.

\section{Dataset and Processing}
\label{sec:data}
We have acquired traces from MazeMap's logging module, covering the period between September 2011 and March 2013. Since the system has been in active development merging logs from different periods required additional effort. Logs contain both explicitly user-initiated (wayfinding, object search and object suggestion) and automatic, periodic client-side, geo-positioning (\emph{geopos}) requests. The number of different requests can seen in Table \ref{tab:req_category}. On one hand, \emph{geopos} requests are initiated every 5-10 seconds (depending on the exact app version), hence the large sample size. On the other hand, individual \emph{geopos} requests are not linked together as there is no permanent user- or session identification implemented in the current system. Moreover, whenever a \emph{geopos} request is initiated outdoors, the system records only that much; the exact position is not saved. Therefore, only aggregate, intra-building statistics are derived from this request type. Since MazeMap is under continuous development, session-tracking and more detailed \emph{geopos} logging features will be available in the future; approval from the institutional review board has already been secured.

\begin{table}[tb]
%\hfill
\centering
\begin{tabular}{|c|c|}
\hline
Type & Sample size\\
\hline \hline
\emph{geopos} & 1301880\\
\hline
wayfinding & 29808\\
\hline
object search & 2294\\
\hline
suggestion & 71\\
\hline
\end{tabular}
\caption{Requests by category}
\label{tab:req_category}
%\hfill
\end{table}

We focus on the Gl\o shaugen campus and filter explicit testing requests from application developers as they would bias our results. We have written Python scripts for data analysis, used Gephi \cite{gephi} for network visualization and QuantumGIS \cite{quantumgis} for map-based visualization. First, we present raw results of spatial and temporal characteristics of requests. In order to draw a human mobility map, we focus on wayfinding requests, and map every step of all turn-by-turn indoor/outdoor routes onto a detailed map of the same granularity. To be able to do this, we submit all wayfinding requests to a backup MazeMap server running the routing algorithm, which returns the corresponding paths. Based on this map, we present an approximation of the aggregated human mobility pattern on campus in Section \ref{sec:geo_results}. We also present the buildings where users spend the most time based on \emph{geopos} requests. Furthermore, we zoom into the largest building (Realfagsb.), and show the same statistics. In Section \ref{sec:social_results}, we only use the sources and destinations of wayfinding requests, and structure them as a weighted graph. We show the trivial and non-trivial logical connections between different locations from a social network perspective; we also find that logical connections between locations show strong dependence on spatial proximity.

Note that the most avid users of the service are freshmen (and new employees), who are not yet familiar with the campus layout. However, their movement patterns are expected to be mostly similar to other students (and employees). Combined with the fact that our logs contain data points from more than 20,000 unique devices, we believe we do not introduce significant sampling bias across the campus population. Furthermore, we assume that users follow the turn-by-turn instructions provided by MazeMap for two reasons. First, maps are constructed from actual floor plans and their connections. Second, MazeMap takes combined indoor/outdoor paths into account; also, the opportunity for cutting corners outdoors is quite limited at this location. We use real building and room names in our results; for detailed context, we refer the reader to the zoomable map of the campus complete with building names at \cite{campus:mazemap}.

\section{Results}
\label{sec:results}

\begin{figure*}[tb]
\centering
%\begin{tabular}{c}
\includegraphics[trim=0cm 0cm 0cm 0cm, clip=false, width=0.95\textwidth]{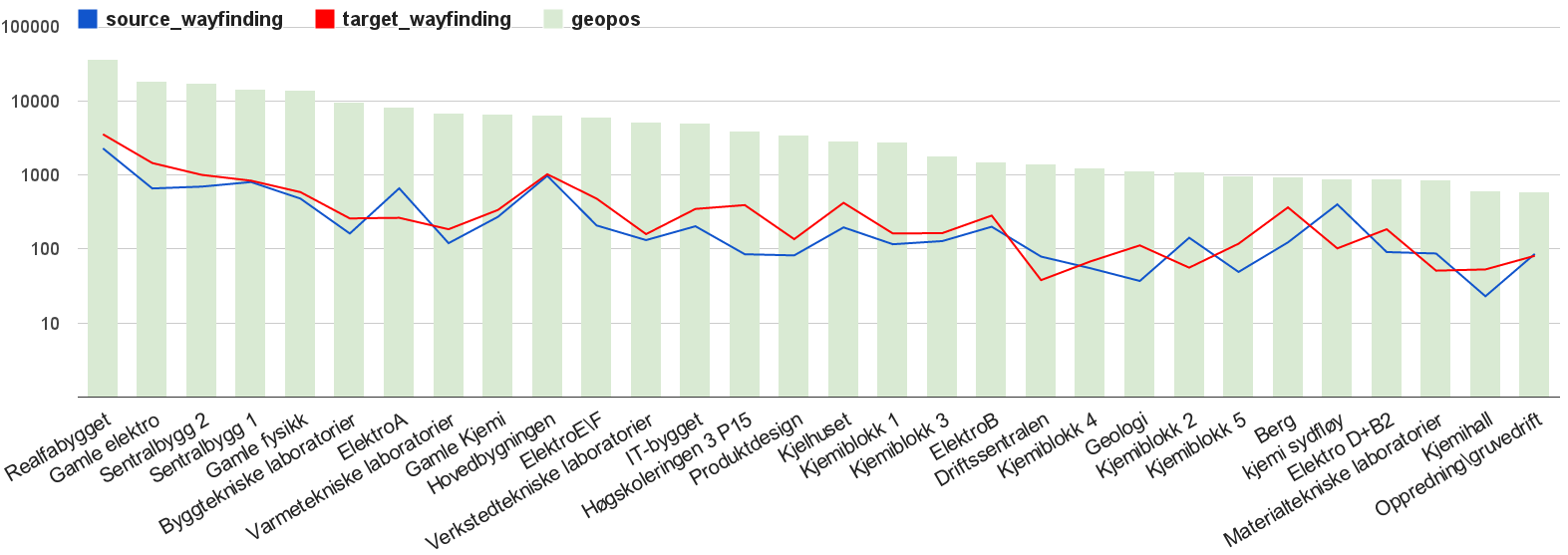}\\
%\end{tabular}
\caption{Number of requests per building (top 30, log scale)}
\label{fig:big_raw}
\end{figure*}

The spatial distribution of \emph{geopos} and \emph{wayfinding} requests is shown in Figure \ref{fig:big_raw} (note the log scale). Since \emph{geopos} requests are sent periodically by active user sessions, they are much larger in volume. They give an approximation on which buildings users  spend the most time at. As expected, buildings with numerous and large lecture halls (Realfagb., Gamle Elektro, Sentralb.) are at the top of the list. The slope of \emph{wayfinding} requests follows a similar pattern, but with certain exceptions, e.g., Hovedb., the main administrative building, is a popular target. It is a less obvious finding that certain buildings exhibit disparity being sources or targets of wayfinding. Berg (general use building) appears several times more as a target, while Elektro A (multiple lecture halls and study rooms) is rather a source. Not shown in the figure is the outdoor area of the campus: a large number of automatic \emph{geopos} requests originate from here, and it is also the largest (albeit aggregated) source of wayfinding queries. Regarding single-room targets, the most popular are large lecture halls (H3, F1 and R1), while smaller halls and the largest cafeteria (Kafe Realfag) are also on the toplist (see Figure \ref{fig:rooms_raw}).

\begin{figure}[tb]
\centering
%\begin{tabular}{c}
\includegraphics[width=0.45\textwidth]{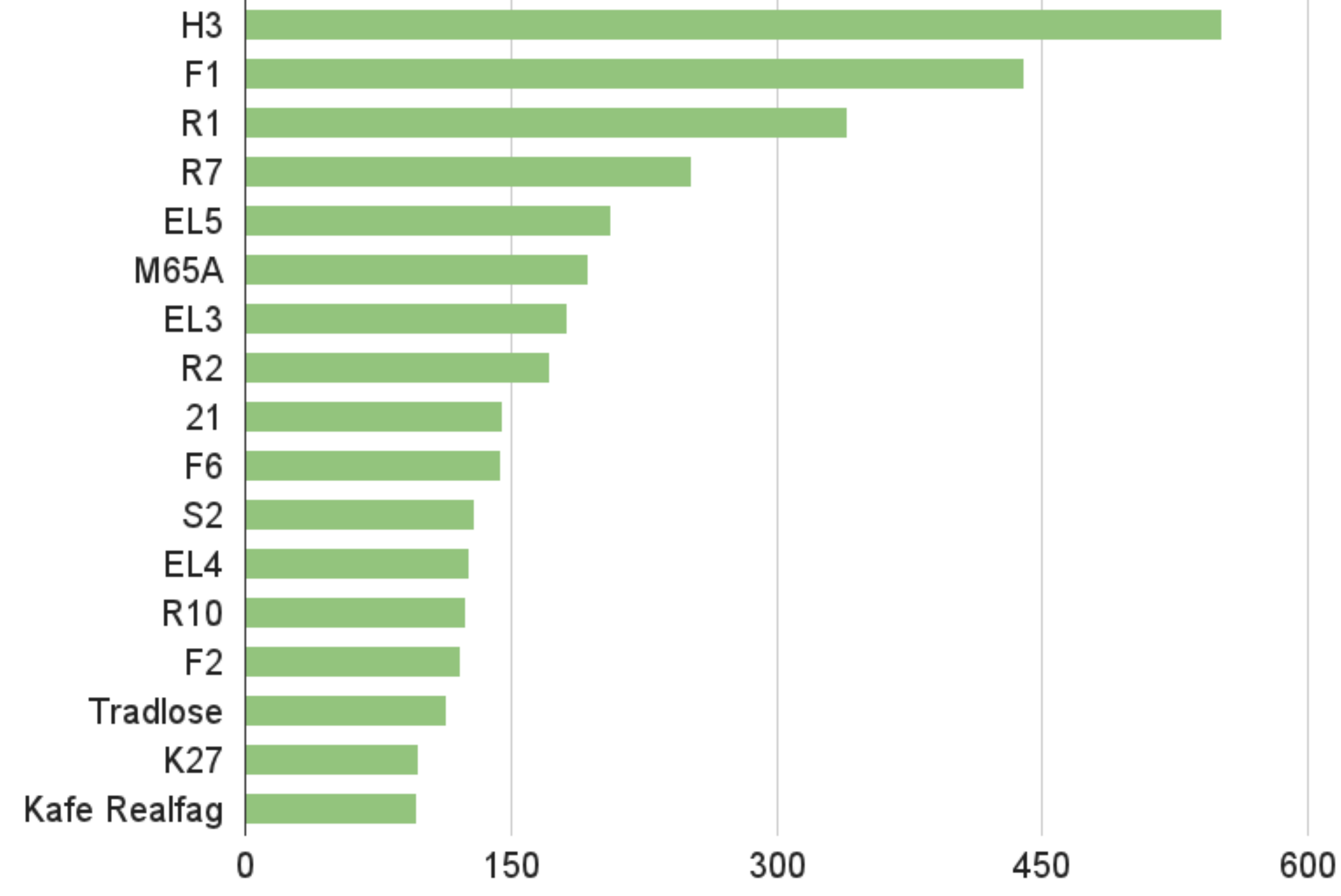}
%\end{tabular}
\caption{Top targets of \emph{wayfinding} requests per room}
\label{fig:rooms_raw}
\end{figure}

Figure \ref{fig:time_raw} shows the temporal distribution of \emph{wayfinding} requests per month over 12 months. The distribution apparently follows a university schedule: most requests are made in the beginning of semesters (August-October and February in Norway), decreasing through the semester (as student get to know the campus better) and bottoming out around exams and holidays (June-July and December-January).

\begin{figure}[tb]
\centering
%\begin{tabular}{c}
\includegraphics[width=0.45\textwidth]{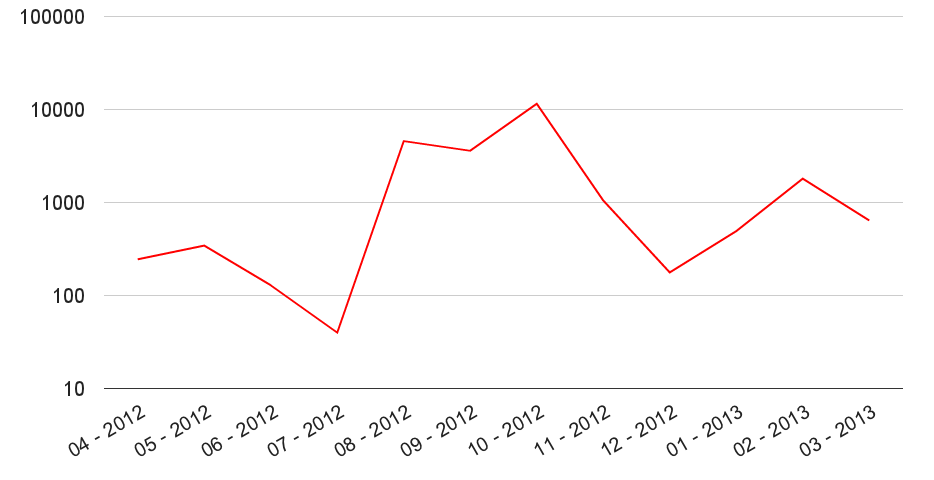}
%\end{tabular}
\caption{Temporal distribution of wayfinding requests (log scale)}
\label{fig:time_raw}
\end{figure}

Regarding object search queries, the most popular POIs are toilets, followed by computer and study rooms. Bus stops and parking lots are less popular, see Figure \ref{fig:objsearch_raw} for the breakdown.

\begin{figure}[tb]
\centering
%\begin{tabular}{c}
\includegraphics[trim=0cm 0cm 0cm 0cm, clip=false, width=0.37\textwidth]{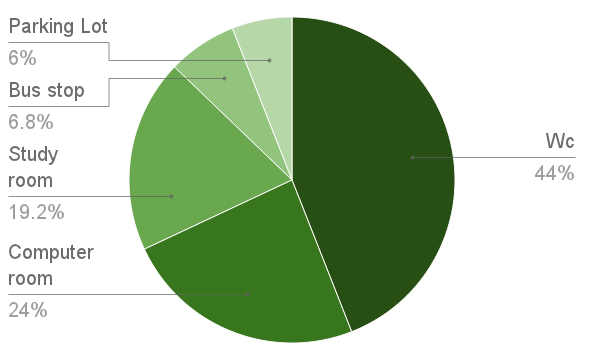}
%\end{tabular}
\caption{Breakdown of requested objects}
\label{fig:objsearch_raw}
\end{figure}

\subsection{Putting statistics on the map}
\label{sec:geo_results}
An interesting interpretation of our dataset is a map of users' movement on campus. Deriving such a ``traffic map'' could be quite important both from scientific and practical points of view, e.g., understanding human mobility or predicting congestion points. We construct such a map based on the 20000 wayfinding requests recorded by MazeMap. Note that this map is an approximation of the real mobility pattern on multiple levels: it uses limited data aggregated over both the user population and time; and it assumes that users actually follow the turn-by-turn instructions provided by the system. 
%Providing more detailed traffic maps showing temporal patterns and individual user trajectories is a challenge addressed in Section \ref{sec:discussion}. 
The aggregated user mobility pattern is shown in Figure \ref{fig:traffic_geo}. The underlying data structure is a weighted, directed graph, where weights denote the frequency of a given path-segment appearing across turn-by-turn routes for all wayfinding requests (directions of edges are omitted for better visibility). Building contours (in red) are layered over the mobility map. A main outdoor pedestrian walkway can be observed running along the campus from S-SE towards N-NW. Several indoor ``highways'' are present, e.g., inside Realfagsb. (with the most large lecture halls) at the south end and Hovedb. (administrative center) at the north end. Almost all narrow corridors and small offices are mapped out by the routes.

\begin{figure}[tb]
\centering
\includegraphics[trim=0cm 0cm 0cm 7cm, clip=true, width=0.5\textwidth
]{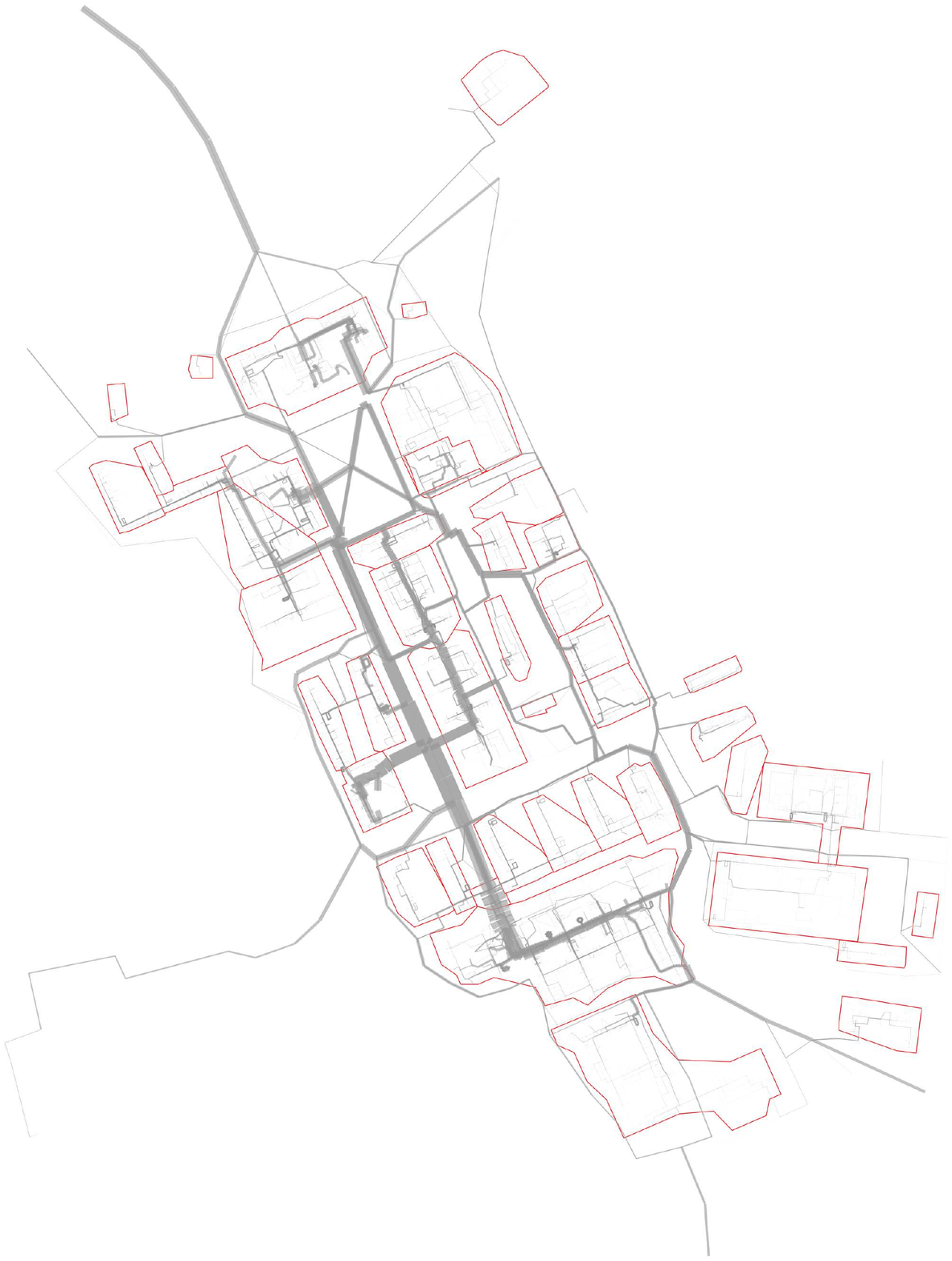}
\caption{Human mobility map: routes calculated from wayfinding requests}
\label{fig:traffic_geo}
\end{figure}

Figure \ref{fig:geopos_geo} shows a building-level map of the campus, colored by the number of automatic \emph{geopos} updates received from the given building (see Figure \ref{fig:big_raw} for exact per building numbers). Small pie charts illustrate whether the building is more a source or target for the users' wayfinding requests. Note that identifying hotspots from \emph{geopos} requests are indeed biased by the actual uptime of apps running on user devices, which in turn is determined by device type, OS type and version, and users explicitly terminating apps. We assume that these effects have been balanced out by the large user population and the long duration of our trace.

\begin{figure}[tb]
\centering
\includegraphics[width=0.435\textwidth]{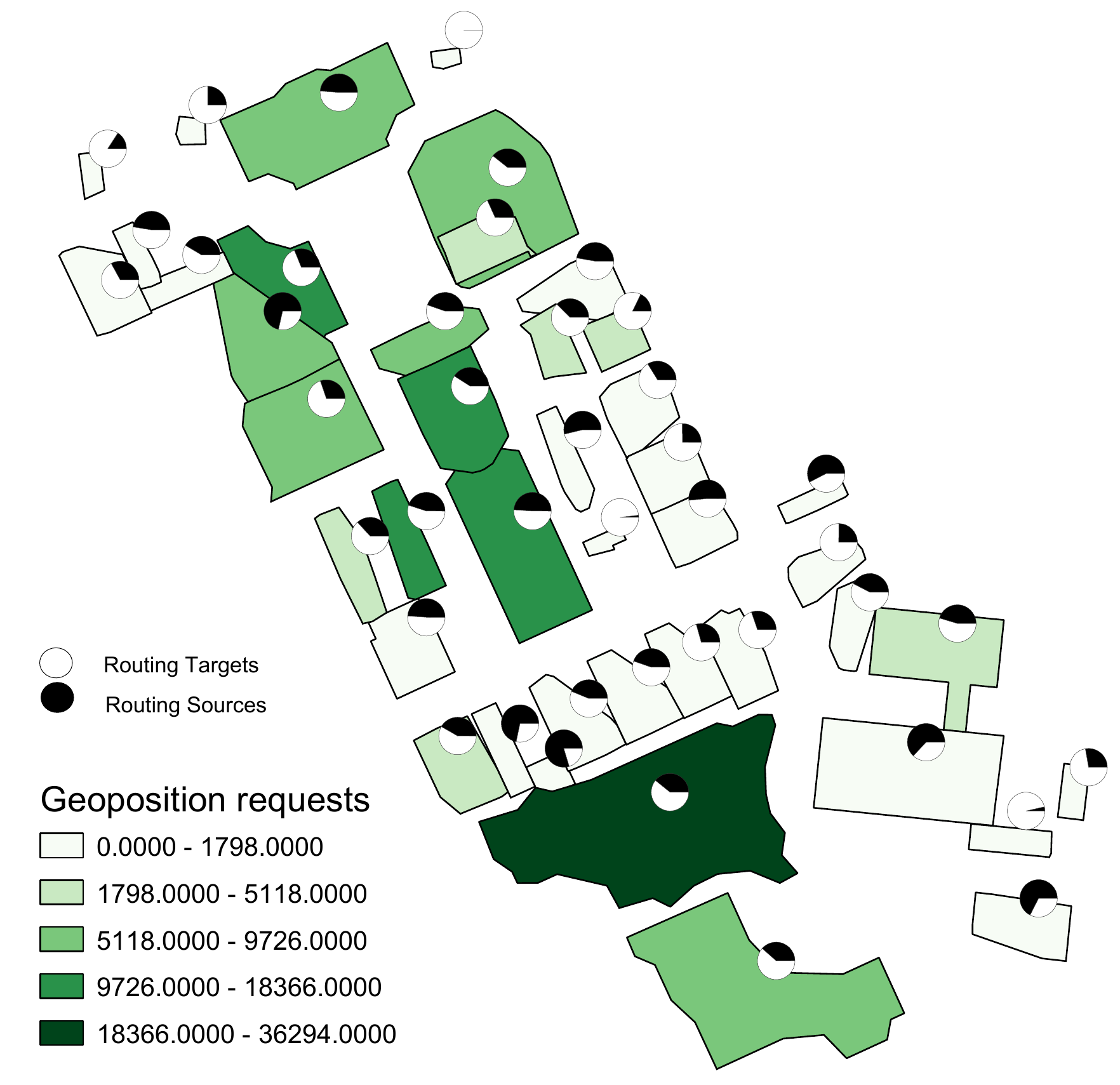}
\caption{Popular buildings: calculated from \emph{geopos} updates}
\label{fig:geopos_geo}
\end{figure}

Statistics from the largest building (Realfagsb.) are visualized in Figure \ref{fig:realfag_geo}. The mobility map is a zoomed in version of Figure \ref{fig:traffic_geo}. Names of rooms which are popular targets of wayfinding requests contain the floor number as the last digit. Obviously, there are numerous other statistics which can be derived for the dataset, both at the building- and the room-level.

\begin{figure*}[tb]
\centering
\includegraphics[width=0.45\textwidth]{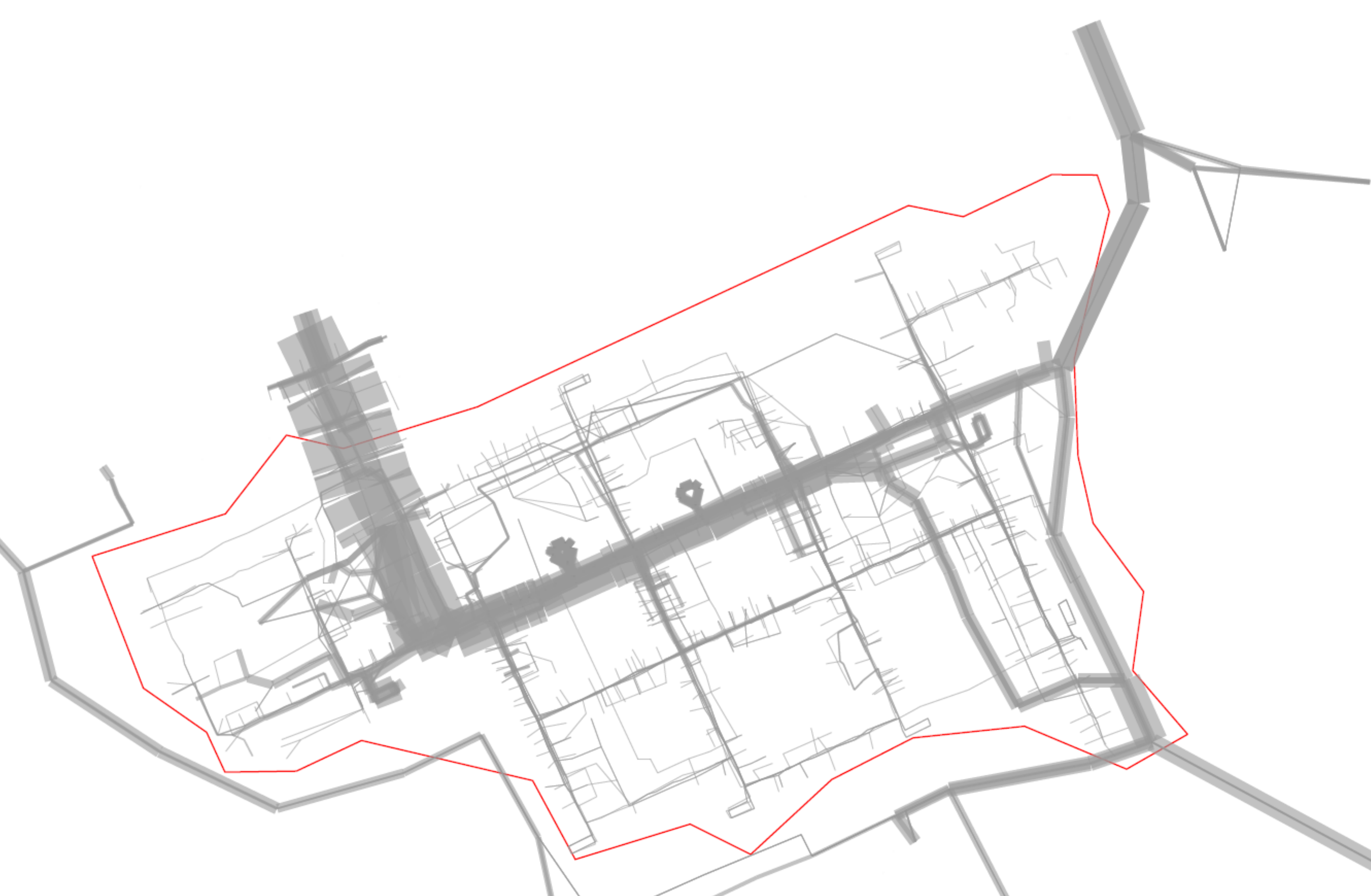}
\hspace{3mm}
\includegraphics[width=0.52\textwidth]{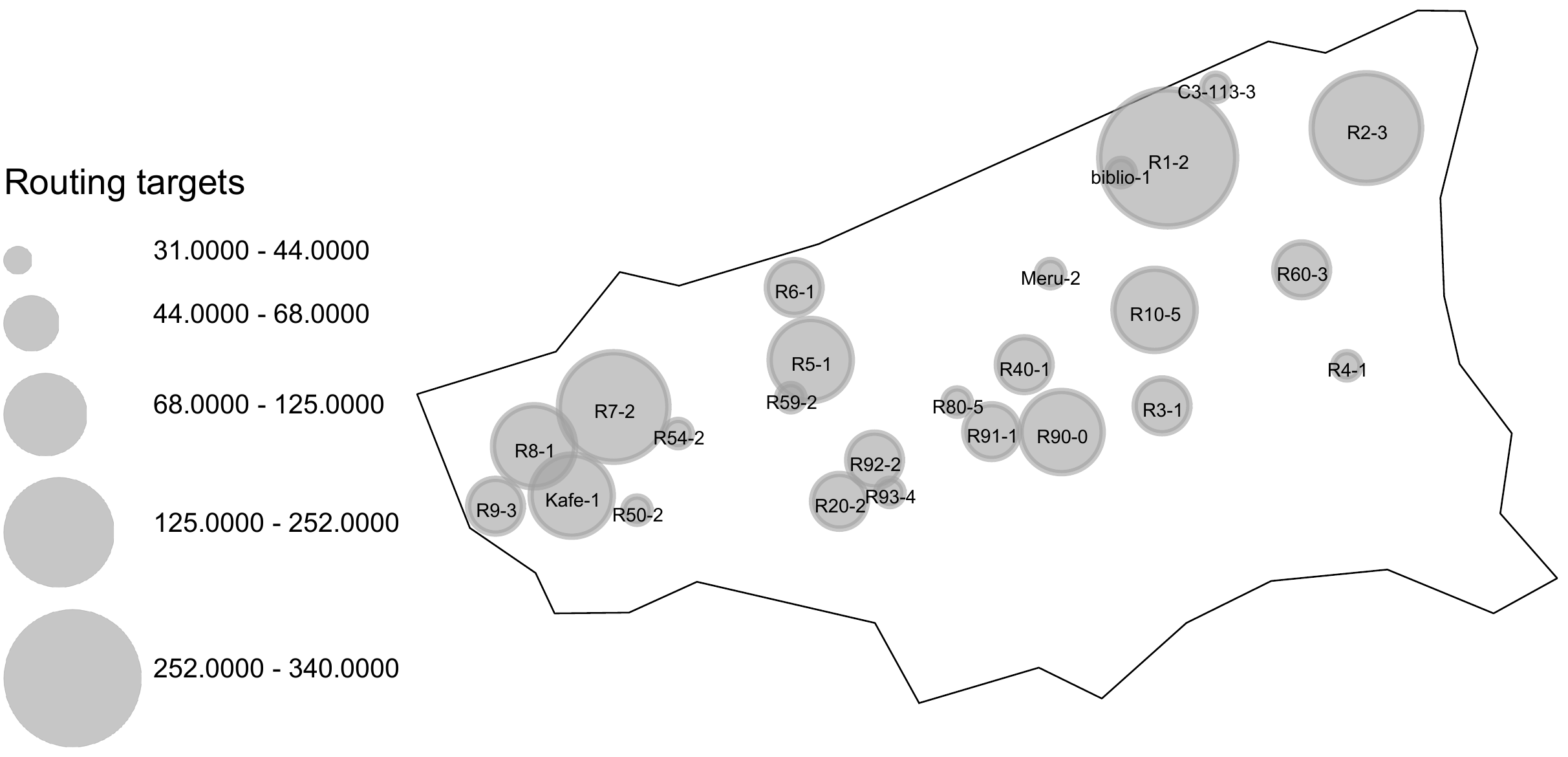}
\caption{Inside the largest building: mobility map (left) and popular target rooms (right)}
\label{fig:realfag_geo}
\end{figure*}

\subsection{Logical connections}
\label{sec:social_results}
If wayfinding requests are viewed as logical connections between rooms, floors and buildings, we can think of locations as nodes in a social network, where the strength of ties is the number of requests between respective nodes (undirected). While floors act as natural communities of rooms, and buildings act as natural communities of floors due to the strong spatial correlation, less trivial strong ties can also be discovered. Figure \ref{fig:openord_social} shows the logical network of floors (nodes are labeled as (floor ID, building ID)), with floors of the same color belonging to the same building, using the OpenOrd layout \cite{openord}. The size of a node corresponds to the weighted node degree, and loop edges denote wayfinding requests within the same floor.  We only show ``strong'' connections above a weight threshold of $10$. We can see that the ground floor ($1.0$) of popular buildings are well-connected; a lot of lecture halls can be found on the ground floor, and NTNU undergrads tend to have lectures covering multiple buildings. However, strong connections do exist between different floors of different buildings, e.g., between the large lecture halls on the ground floor of Gamle Elektro $(1.0,1)$, and the third floor of Hovedb. $(3.0,2)$, where the financial department is located.

\begin{figure}[tb]
\centering
\includegraphics[width=0.35\textwidth]{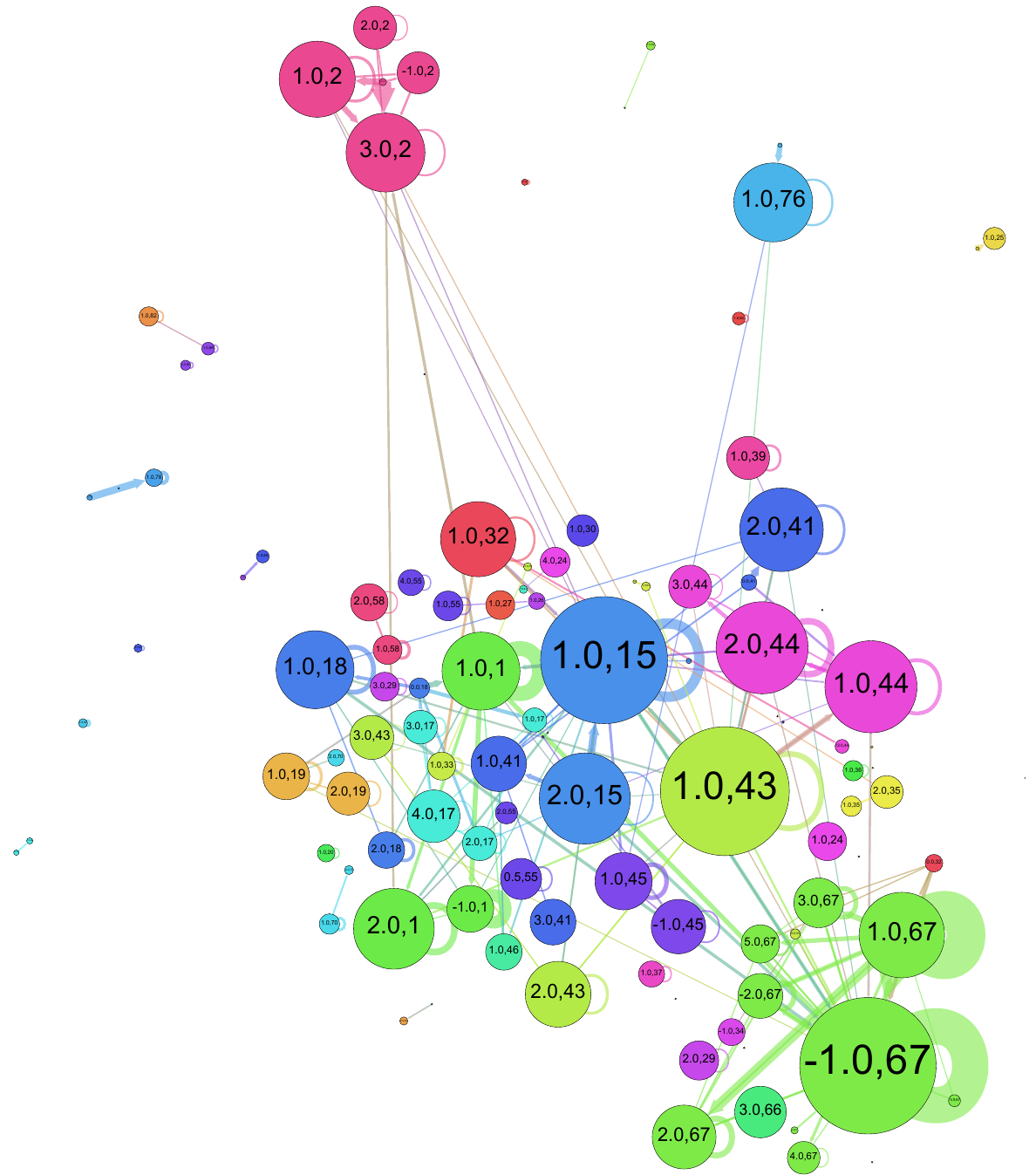}
\caption{Logical network of floors based on wayfinding requests. Nodes: (floorID, buildingID). Color codes represent buildings.}
\label{fig:openord_social}
\end{figure}

Based on the logical network above, we suspect that logical connections are strongly correlated with spatial proximity. In Figure \ref{fig:cdf_social}, we plot the cumulative distribution function (CDF) of weighted logical connections with respect to geographical distance between their endpoints (simplified as the distance between buildings). It is easy to see that close to 70\% of the weight of connections situated indoors (zero distance), and $\approx$ 90\% of them correspond to a geographical distance less than 500 meters. The ``traffic map'' in Figure \ref{fig:traffic_geo} hints at the opposite; however, indoor-only paths consist of much less path segments therefore being less visible on the map.

\begin{figure}[tb]
\centering
\includegraphics[width=0.45\textwidth]{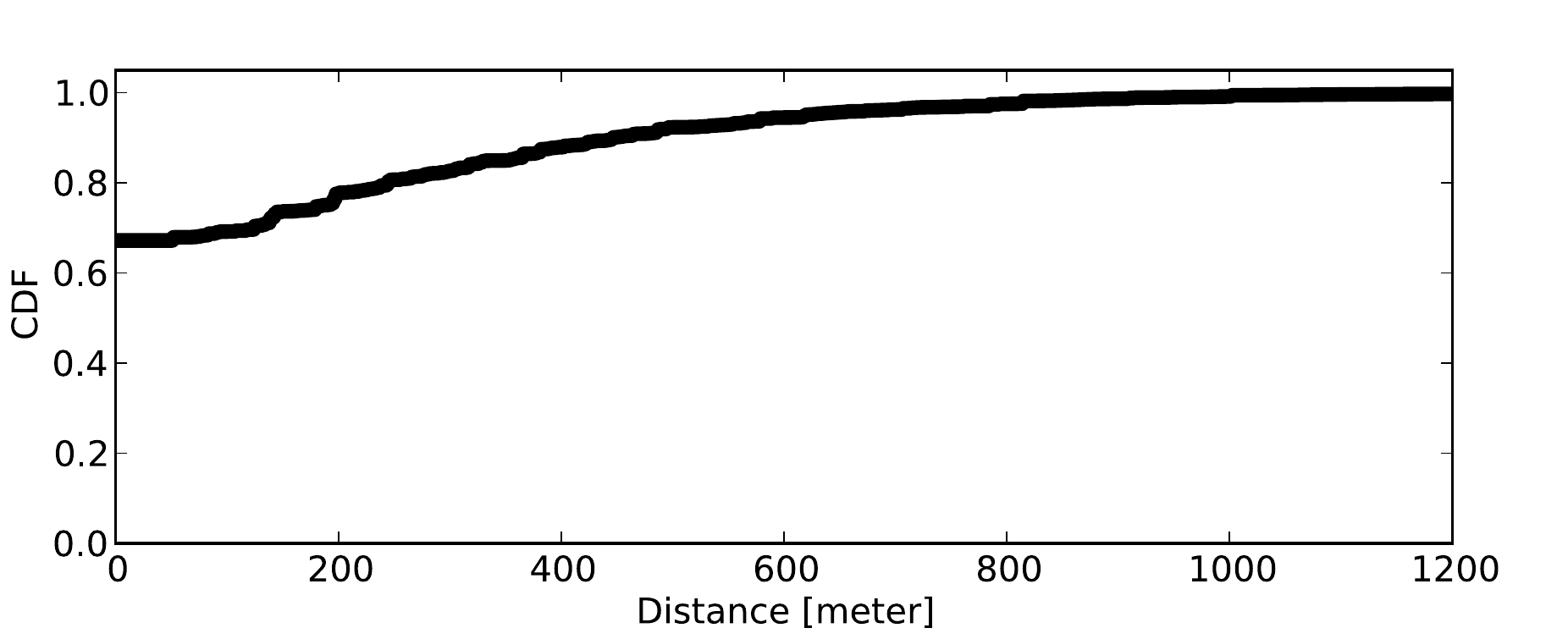}
\caption{CDF: geographical distance of weighted logical connections}
\label{fig:cdf_social}
\end{figure}

Please find all of our maps in zoomable vector graphics format at \cite{svg_dropbox}.

\section{Discussion}
\label{sec:discussion}
Results presented in Section \ref{sec:results} form an interesting campus case study. However, IPSs could potentially offer much more data, and enable various location-based services at numerous other venues.

\textbf{Potential usage on-campus.} Managing location resources at the university-level is not easy: evidence from an IPS providing data on actual user movement, distances to cover in breaks between classes and potential bottleneck staircases and corridors can make the outcome closer to optimal. Coordination among faculties and departments concerning meeting and lecture rooms can also be improved.  

\textbf{Potential for other scenarios: shopping malls, hospitals and emergency preparedness.} A venue for IPS-based intelligence with huge business potential is the shopping mall; here, being able to easily get mobility and popularity data with different granularity (room, floor, wing, building, venue) could be extremely helpful \cite{indoor_shop_shopkick}. Tailor-made, location-specific advertising could be realized. Studying human traffic patterns (see Figure \ref{fig:traffic_geo}), mall managers can validate the position of or relocate their flagship stores, get informed on where to plug in a novelty store, or work out an informed pricing plan for tenants. In addition, studying the logical connections among stores (in the flavor of Figure \ref{fig:openord_social}, studying tie strengths, communities, etc.) can shed light on non-trivial shopping patterns, creating the possibility for a profitable coupon program and product bundling encompassing multiple stores of different profiles. Furthermore, evidence from deployed IPSs can be used when planning, building and operating new shopping malls, office buildings or hospitals. Specifically in hospitals, it is essential to keep track of where both patients, staff and equipment are, and support both wayfinding (for patients), planning (of logistics), and self-coordination (of doctors and nurses) as unexpected events occur \cite{Nossum:2009}. Finally, regarding safety, IPSs could enable a higher level of emergency preparedness at indoor venues. Response teams can be sent to an exact location, and evacuation routes could be adjusted on the fly; this approach could be also utilized in large underground complexes and tunnels, where the GPS system has no use.
 
\textbf{Human mobility research.} Understanding human mobility patterns is of great importance from scientific, engineering (how to build more efficient systems for people) and business angles. Owing to large mobile phone traces, we have a pretty good understanding of people's movements on a macro scale \cite{gonzalez2008understanding}. However, since urban population spends the majority of its lifetime indoors, understanding micro-scale mobility is essential; once equipped with an individual user tracking feature, IPSs can help tremendously in this area.

\textbf{Challenges.} All potential services and usage scenarios mentioned above can only be built on carefully designed and implemented IPSs, posing technical challenges. The two most important requirements for IPS are precision and scalability (location update frequency, peak load, large venues). Data quality of the premises-data are very important. Another important aspect is the integration with GPS-based systems and other, more fine-grained location services. In addition to the technical, there is a need for deep understanding of location privacy \cite{beresford2003location} and users' valuation of privacy \cite{Kofod:2009}, and data handling best practices such as with network measurement data \cite{coull2007playing}. It is clear that further development and potential success of IPS technology and location-based services depend on a cooperative effort from researchers, engineers, economists and lawmakers.
%\cite{phelps2000privacy}

\section{Conclusion}
\label{sec:conclusion}
In this paper we provided a glimpse into data available from a hybrid indoor/outdoor positioning and navigation system called MazeMap. We showed possible interpretations of these data, including finding potential bottlenecks and hotspots with the help of mapping user mobility patterns; reflecting on spatial (room, building and campus-level) and temporal characteristics of user requests; and discovering spatio-logical connections between locations in different buildings. Justified by our case study, we presented a brief outlook on the potential of indoor positioning systems and location-based services enabled by them. We believe that such systems and services have a bright future with regard to both data-driven research, engineering and business applications.

\bibliographystyle{abbrv}
\bibliography{mazemap_permoby}

\begin{thebibliography}{10}

\bibitem{indoor_personalized_museum}
{American Museum of Natural History}.
\newblock \url{http://www.amnh.org/apps/explorer}. Last accessed: Nov 2013.

\bibitem{Andresen:2007}
S.~Andresen, J.~Krogstie, and T.~Jelle.
\newblock Lab and research activities at wireless trondheim.
\newblock In {\em Proc. of the 4th IEEE International Symposium on Wireless
  Communication Systems (ISWCS'07)}, 2007.

\bibitem{svg_dropbox}
{Appendix: maps in SVG format}.
\newblock
  \url{https://dl.dropboxusercontent.com/u/3806223/appendix_mazemap_permoby.zip}.

\bibitem{indoor_apple}
{Apple acquires indoor location company WiFiSLAM}.
\newblock
  \url{http://blogs.wsj.com/digits/2013/03/23/apple-acquires-indoor-location\\-company-wifislam/}.
  Last accessed: Nov 2013.

\bibitem{beresford2003location}
A.~R. Beresford and F.~Stajano.
\newblock Location privacy in pervasive computing.
\newblock {\em Pervasive Computing, IEEE}, 2(1):46--55, 2003.

\bibitem{indoor_shop_bestbuy}
{BestBuy}.
\newblock
  \url{http://www.mobilecommercedaily.com/best-buy-geofences-locations-to-deliver\\-targeted-mobile-offers}.
  Last accessed: Nov 2013.

\bibitem{indoor_venue_hospital}
{Boston Children's Hospital}.
\newblock
  \url{http://childrenshospital.org/patientsfamilies/Site1393/mainpageS1393P474.html}.
  Last accessed: Nov 2013.

\bibitem{indoor_chip_broadcom}
{Broadcom GNSS product features}.
\newblock \url{http://www.broadcom.com/products/features/GNSS.php}. Last
  accessed: Nov 2013.

\bibitem{campusguide_press}
{CampusGuide: 'GPS' for Indoor Use}.
\newblock \url{http://www.sciencedaily.com/releases/2013/02/130206093907.htm}.
  Last accessed: Nov 2013.

\bibitem{coull2007playing}
S.~Coull, C.~Wright, F.~Monrose, M.~Collins, M.~K. Reiter, et~al.
\newblock Playing devil's advocate: Inferring sensitive information from
  anonymized network traces.
\newblock In {\em Proc. of the Network and Distributed System Security
  Symposium}, 2007.

\bibitem{indoor_forbes}
{Forbes: Indoor venues are the next frontier}.
\newblock
  \url{http://www.forbes.com/sites/forrester/2013/01/23/indoor-venues-are-the-next-frontier-for\\-location-based-services/}.
  Last accessed: Nov 2013.

\bibitem{gephi}
{Gephi website}.
\newblock \url{https://gephi.org/}. Last accessed: Nov 2013.

\bibitem{gonzalez2008understanding}
M.~C. Gonzalez, C.~A. Hidalgo, and A.-L. Barabasi.
\newblock Understanding individual human mobility patterns.
\newblock {\em Nature}, 453(7196):779--782, 2008.

\bibitem{indoor_map_google}
{Google Maps Indoor}.
\newblock \url{http://maps.google.com/help/maps/indoormaps/}. Last accessed:
  Nov 2013.

\bibitem{gps_report}
{GPS Market Forecast}.
\newblock
  \url{http://www.reportlinker.com/p0936054-summary/GPS-Market-Forecast-to.html}.
  Last accessed: Nov 2013.

\bibitem{ips_survey}
Y.~Gu, A.~Lo, and I.~Niemegeers.
\newblock A survey of indoor positioning systems for wireless personal
  networks.
\newblock {\em Communications Surveys \& Tutorials, IEEE}, 11(1):13--32, 2009.

\bibitem{indoor_people}
P.~L. Jenkins, T.~J. Phillips, E.~J. Mulberg, and S.~P. Hui.
\newblock Activity patterns of californians: use of and proximity to indoor
  pollutant sources.
\newblock {\em Atmospheric Environment. Part A.}, 26(12):2141--2148, 1992.

\bibitem{Kofod:2009}
A.~Kofod-Petersen, P.~A. Grans\ae~ter, and J.~Krogstie.
\newblock An empirical investigation of attitude towards location-aware social
  network service.
\newblock {\em International Journal of Mobile Communications}, 8:53--70, 2009.

\bibitem{Krogstie:2012}
J.~Krogstie.
\newblock Bridging research and innovation by applying living labs for design
  science research.
\newblock In {\em Proc. of SCIS}, pages 161--176, 2012.

\bibitem{trilateration}
B.~Li, J.~Salter, A.~G. Dempster, and C.~Rizos.
\newblock Indoor positioning techniques based on wireless lan.
\newblock In {\em Proc. of IEEE Int. Conf. on Wireless Broadband and Ultra
  Wideband Communications}, 2007.

\bibitem{openord}
S.~Martin, W.~M. Brown, R.~Klavans, and K.~W. Boyack.
\newblock Openord: an open-source toolbox for large graph layout.
\newblock In {\em Proc. of IS\&T/SPIE Electronic Imaging}, 2011.

\bibitem{campusguide}
{MazeMap website}.
\newblock \url{http://mazemap.com}. Last accessed: Nov 2013.

\bibitem{indoor_map_nokia}
{Nokia leads the way with indoor mapping}.
\newblock
  \url{http://conversations.nokia.com/2012/07/16/nokia-leads-the-way-with-indoor-mapping/}.
  Last accessed: Nov 2013.

\bibitem{Nossum:2009}
A.~Nossum and J.~Krogstie.
\newblock Integrated quality of models and quality of maps.
\newblock In {\em Proc. of EMMSAD}, pages 264--276, 2009.

\bibitem{campus:mazemap}
{NTNU campus map at MazeMap}.
\newblock \url{http://use.mazemap.com}.

\bibitem{spatial:2013}
A.~Petrenko, S.~Bell, K.~Stanley, W.~Qian, A.~Sizo, and D.~Knowles.
\newblock Human spatial behavior, sensor informatics, and disaggregate data.
\newblock In {\em Proc. of Spatial Information Theory 2013}, pages 224--242.
  2013.

\bibitem{quantumgis}
{QuantumGIS website}.
\newblock \url{http://www.qgis.org/}. Last accessed: Nov 2013.

\bibitem{indoor_shop_shopkick}
{ShopKick says it's now profitable}.
\newblock
  \url{http://techcrunch.com/2013/01/16/shopkick-says-its-now-profitable-with-its\\-shopping-app-adding-200m-in-sales-for\\-target-best-buy-and-other-partners/}.
  Last accessed: Nov 2013.

\end{thebibliography}
\end{document}